\documentclass[prl,twocolumn]{revtex4-1}
\usepackage{amsfonts,amssymb,amsmath,bm}
\usepackage{graphicx,epsfig,color}

\begin{document}

\title{Universal profile of the vortex condensate in two-dimensional turbulence}
\author{Jason Laurie$^{1}$, Guido Boffetta$^2$, Gregory Falkovich$^{1,3}$, Igor Kolokolov$^{4,5}$, and Vladimir Lebedev$^{4,5}$}

\affiliation{
$^1$Department of Physics of Complex Systems, Weizmann Institute of Science, Rehovot, 76100, Israel \\
$^2$Dipartmento di Fisica and INFN, Universit\`{a} di Torino, via P. Giuria 1, 10125 Torino, Italy \\
$^3$Institute for Information Transmission Problems, Moscow, 127994, Russia; \\
$^4$Landau Institute for Theoretical Physics, Kosygina 2, Moscow, 119334, Russia; \\
$^5$Moscow Institute of Physics and Technology, Dolgoprudny, Moscow, 141700, Russia}

\date{\today}

\begin{abstract}

An inverse turbulent cascade in a restricted two-dimensional periodic domain
leads to the creation of condensate -- a pair of coherent system-size vortices. We perform extensive numerical simulations of this system and carry on detailed theoretical analysis based upon momentum and energy exchanges between the turbulence fluctuations and the mean coherent condensate (vortices). The theory predicts the vortex profile and amplitude which
perfectly agree with the numerical data.

\end{abstract}


\maketitle

From both a fundamental and practical perspective, a central problem of
turbulence theory is the understanding and the description of the
interaction of turbulence fluctuations with a mean (coherent) flow
\cite{townsend80}. Even at the basic level of energy and momentum budget such interaction is quite non-trivial: we expect energy to go from the mean flow to turbulence in the fully three-dimensional case while it can go from turbulence to the mean flow in the two-dimensional (2D) case \cite{BE12} or in fluid layers \cite{Xia}. At present, there is no unified conceptual framework
to address this problem. The cases most studied  are wall bounded flows in channels or pipes, for which experiments and numerical studies have been done for over a century. Despite this, even basic problems such as to determine at which mean velocity turbulent fluctuations are sustained is still object of
intense investigations \cite{avila2011}; nor is there any consistent theory for the mean profile with turbulence, so that even the celebrated logarithmic law is a subject of controversy \cite{deb}. Here, we consider 2D turbulence in a restricted box where large-scale coherent structures are generated from small-scale fluctuations excited by pumping. This process occurs because the 2D Navier-Stokes equation favors energy transfer to larger scales \cite{BE12,Xia,67Kra,68Lei,69Bat} -- a phenomenon known as the inverse cascade.

Already, the first experiments on 2D turbulence \cite{Som} have
shown that in a finite system with small bottom friction, the inverse
cascade leads to the formation of coherent vortices. Subsequent simulations
\cite{Colm} and experiments \cite{Shats} demonstrated that these vortices
have well-defined mean vorticity profiles with a power-law radial decay. In this paper, we present the results of new extensive simulations of 2D turbulence in a periodic box. We analyze the structure of the coherent vortices in the presence of a friction term. Dealing with a statistically steady state enables us to collect extensive statistics. We propose a new theoretical framework for the analysis of turbulence-flow interaction, explaining the numerical results and giving new insight in the coherent vortices formation and structure.

The starting point for both the theory and the numerical simulations is the forced 2D Navier-Stokes equation for the 2D velocity field $\bm v$ with linear bottom friction,
 \begin{equation}
 \partial_t{\bm v}+\alpha {\bm v}
 +({\bm v}\cdot\nabla){\bm v}
 =-\nabla p+\nu\Delta\bm v+{\bm f},
 \label{Navier-Stokes}
 \end{equation}
where $\alpha$ is the friction coefficient, $\bm f$ is an external force (per
unit mass) exciting the turbulence, and $\nu$ is the kinematic viscosity. The
force $\bm f(t,x,y)$ is assumed to be a random function with homogeneous
statistics, with a forcing correlation time small enough and a correlation
length much less than the system size $L$. The coefficient $\alpha$ is
assumed to be small comparing to the inverse turnover time of the system-size vortices, $\alpha^3 \ll \epsilon/L^2$, where $\epsilon=\langle \bm
f\cdot \bm v \rangle$ is the energy production rate (per unit mass). The
angular brackets here and below designate temporal averaging.

\begin{figure}[h!]
\begin{center}
\includegraphics[width=0.8\columnwidth]{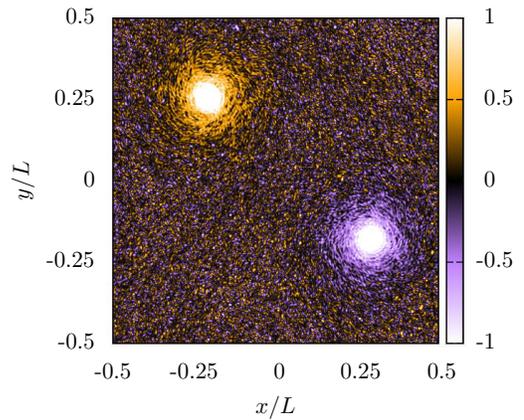}
\caption{(color online) Plot of the total vorticity
during the condensate regime of simulation B. \label{fig:1}}
\end{center}
\end{figure}

In our simulations, we use a periodic square box of size $L=2\pi$ so that the Fourier grid spacing is $dk =2\pi/L=1$. We numerically solve the 2D Navier-Stokes equation with linear friction \eqref{Navier-Stokes} using a
pseudo-spectral spatial method, fully de-aliased by the 2/3rds rule and time
stepped by a second order Runge-Kutta scheme. The spatial resolution is
$512\times 512$. The external forcing acts in Fourier space in an annulus
of width $3 dk=3$  centered around the forcing wave number $k_f=100$ with a constant amplitude of $0.1$. We replace the small-scale viscous dissipation term in \eqref{Navier-Stokes} by hyper-viscosity: $\nu(-\Delta)^8{\bm v}$  with $\nu=5\times 10^{-35}$ to provide an extended inertial range in Fourier space and to better reveal any universality of the vortex condensate. In all simulations, the forcing and the hyper-viscous term (including coefficients) are identical.

We perform three sets of simulations, with the only difference being the
linear friction coefficient: $\alpha=1.1\times 10^{-4}$ (A),
$6.4\times 10^{-5}$ (B) and $3.2\times10^{-5}$ (C), which results in
slightly different inverse energy fluxes $\epsilon =  3.47\times 10^{-4}$ (A),
$3.57\times 10^{-4}$ (B) and $3.47\times 10^{-4}$ (C). Each simulation is run until the system reaches a non-equilibrium stationary state through the balance of the forcing and linear friction term, observed by the time stationarity of the total kinetic energy $E=(1/2)\int dx\,dy\,{\bm v}^2$. Once stationary, we output data at every large eddy turnover time estimated by assuming that the total energy is dominated by the condensate at the largest scale for $4\times 10^4$ large eddy turnover times.  A typical snapshot of the vorticity field in the stationary state is plotted in Fig. \ref{fig:1}. For disentangling the mean flow from the turbulence, it is crucial  to locate the vortex center and then to follow it as the vortex pair wanders in space. For each time frame, we locate the center of the positive (vorticity) vortex by determining the global maximum of the vorticity and then computing the center of mass of the vorticity in a box of $8\times 8$ grid points around the extremum. Subsequently, we shift the domain at every step so that the vortex center is located at the origin.

\begin{figure}
\begin{center}
\includegraphics[width=0.8\columnwidth]{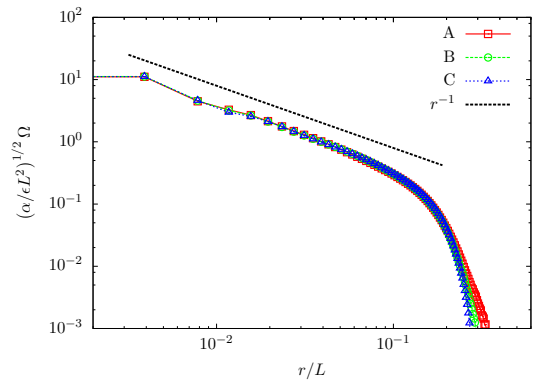}
\caption{(color online) Radial profile of the mean vorticity $\Omega$ normalized by $(\epsilon L^2/\alpha)^{1/2}$.  The straight black dashed line corresponds to a radial profile $\propto r^{-1}$. \label{fig:2}}
\end{center}
\end{figure}

\begin{figure}
\begin{center}
\includegraphics[width=0.8\columnwidth]{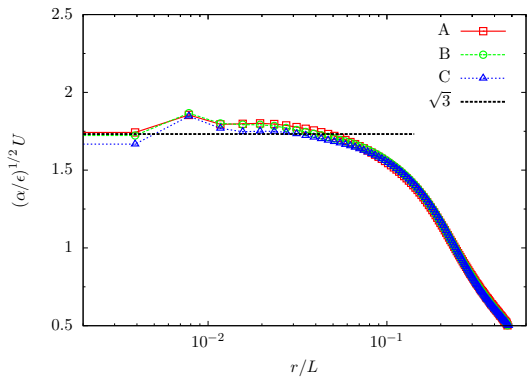}
\caption{(color online) Radial profile  of the mean polar velocity
$U$ normalized by $(\epsilon/\alpha)^{1/2}$ plotted in log-lin coordinates.
The straight horizontal black dashed line corresponds to
$(\alpha/\epsilon)^{1/2}U=\sqrt{3}$, Eq.~\eqref{Uni}.
\label{fig:3}}
\end{center}
\end{figure}

\begin{figure}
\begin{center}
\includegraphics[width=0.8\columnwidth]{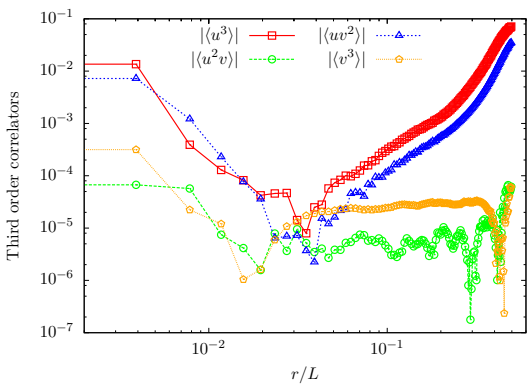}
\caption{(color online) Radial profiles of the absolute values of the third order moments, $\left\langle u^3\right\rangle$, $\left\langle u^2v\right\rangle$, $\left\langle uv^2\right\rangle$, and $\left\langle v^3\right\rangle$ for simulation C.  One observes an additional smallness of all odd in $v$ moments compared to their even in $v$ counterparts. \label{fig:4}}
\end{center}
\end{figure}

\begin{figure}
\begin{center}
\includegraphics[width=0.8\columnwidth]{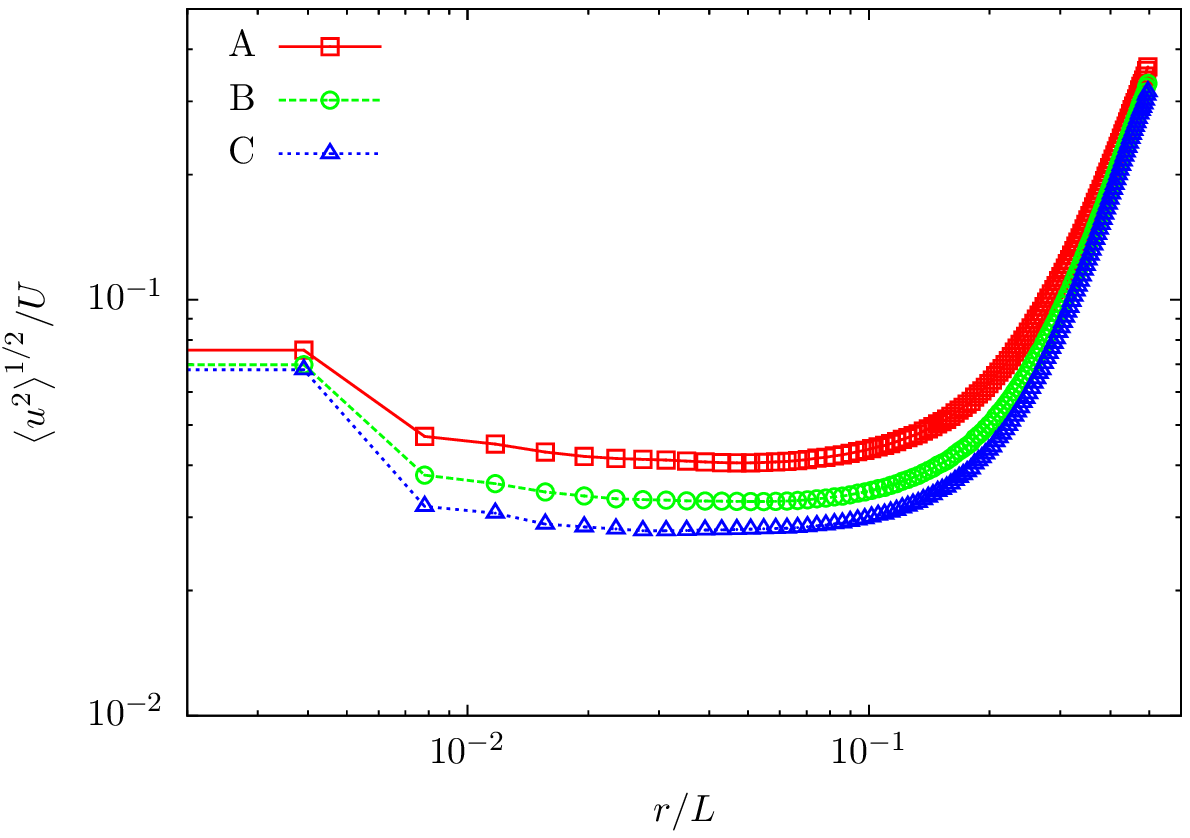}
\caption{(color online) Radial profile  of the rms fluctuating polar
velocity $u$ normalized by the mean polar velocity $U$.  \label{fig:5}}
\end{center}
\end{figure}

\begin{figure}
\begin{center}
\includegraphics[width=0.8\columnwidth]{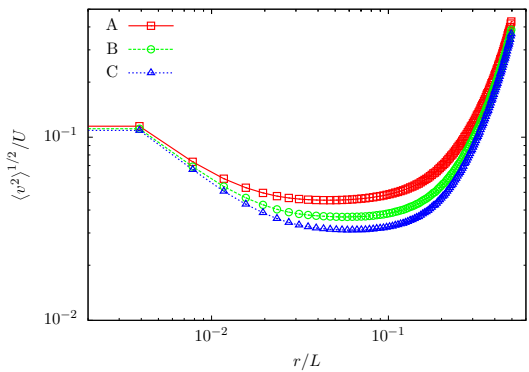}
\caption{(color online) Radial profile  of the rms fluctuating
radial velocity $v$ normalized by the mean polar velocity $U$. \label{fig:6}}
\end{center}
\end{figure}

\begin{figure}
\begin{center}
\includegraphics[width=0.8\columnwidth]{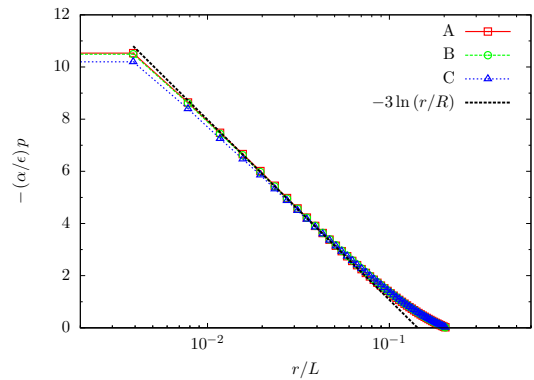}
\caption{(color online) Radial profile of the pressure  $p$ normalized by $-(\epsilon/\alpha)$. The numerical data are compared to the theoretical prediction of Eq.~\eqref{pressure} using the fitting parameter $R/L=0.143$. 
\label{fig:7}}
\end{center}
\end{figure}

The decomposition into the mean and fluctuating components is made by
performing a temporal average over all time frames of the spatially centered
vortex to filter out the zero-mean fluctuations and to subsequently obtain the
mean vorticity distribution. The respective fluctuations are then acquired by
subtracting the mean flow from the original vorticity distribution. We double
the statistics by applying the same method to the other (negative vorticity)
vortex in the condensate after the required vorticity-velocity symmetry
transformations to permit us to change sign of the vorticity. Results of the temporal averaging for simulations with different linear friction coefficients are presented in Figs.~\ref{fig:2}-\ref{fig:7}. The amplitude of the final condensate apparently scales as $\alpha^{-1/2}$.

The mean velocity profile inside the vortex is highly isotropic.  The vortex interior can be separated into the vortex core and the region outside the core where the average velocity profile reveals some universal scaling properties.  We focus on this universal behavior.

Let us now provide some basic theoretical analysis. We introduce polar coordinates in the reference system with the origin at the vortex center:
$r$ is the distance from the vortex center and $\varphi$ is the corresponding polar angle. Based upon numerical simulations and experiments, we assume that the vortex is isotropic that it can be described in terms of the average (over time) polar velocity $U$, which is a function of $r$. The same is assumed for the average vorticity $\Omega =U/r+\partial_r U$. Taking the curl of the 2D Navier-Stokes equation (\ref{Navier-Stokes}), neglecting the viscous term (assumed to be small for scales larger than the pumping length) and decomposing the mean flow from the fluctuations, one obtains
 \begin{eqnarray}
 \alpha \Omega + \frac{1}{r} \partial_r
 \left(r\langle v \omega \rangle\right) =0,
 \label{basic2} \\
 \partial_t\omega +\frac{U}{r}\partial_\varphi \omega
 +v \partial_r\Omega +\alpha \omega 
 \nonumber\\
 =-[v\partial_r+(u/r)\partial_\varphi] \omega
 - \alpha \Omega
 +\mathrm{curl}\, {\bm f},
 \label{flu1}
 \end{eqnarray}
where $v$ is the radial component of the fluctuating velocity, $u$ is its polar component and $\omega$ is the fluctuating vorticity.

An attempt to construct a theory explaining the power-law profile $\Omega\propto r^{-a}$ was made in \cite{Chert}. It was based on the existence of power-law zero modes of $\omega$ on the background of the power-law averaged profile $\Omega$. Assuming that the zero modes give the main contribution to the mean vorticity flux $\langle v \omega \rangle$ and using perturbation theory (over non-linear interaction) one can relate  $a$ to the power-law scaling of the hypothetical leading contribution to $\langle v \omega \rangle$. Equating scaling exponents of both parts of (\ref{basic2}), one finds $a=5/4$
\cite{Chert}, that does not contradict the  results of \cite{Colm,Shats}. Our
data, with higher resolution and increased statistics, suggests however
that $a\approx 1$, see Fig. \ref{fig:2}. This is even more clear from the mean velocity profile, plotted in Fig. \ref{fig:3}, which demonstrates that $U$ is $r$-independent inside the vortex, in accordance with the dependence $\Omega\propto r^{-1}$.

To explain the discrepancy between the zero-mode prediction and the actual profile, here we note that the zero modes must give an anomalously small contribution to the average $\langle v \omega \rangle$. This follows from symmetry consideration. Indeed, consider Eq.~(\ref{flu1}). If, as assumed in \cite{Chert}, that the pumping term on the right-hand side of (\ref{flu1}) can be neglected for large-scale motions (i.e. when the characteristic scale $r$ exceeds the pumping correlation length), then multiplying (\ref{flu1}) by $\omega^n$ and averaging over time one obtains
 \begin{eqnarray}
 \langle v \omega^n \rangle \partial_r\Omega
 +\frac{1}{(n+1)r} \partial_r \langle r v \omega^{n+1}\rangle
 \nonumber \\
 +\alpha \langle \omega^{n+1} \rangle
 +\alpha \langle \omega^n \rangle \Omega=0,
 \nonumber
 \end{eqnarray}
where we have used isotropy. From the set of relations for different $n$, it follows that the large-scale contributions to $\langle v \omega^n \rangle$ are proportional to $\alpha$ and tend to zero as $\alpha\to0$. The same is valid for other averages odd in $v$. On a deeper level this follows from time-reversibility of the Euler equation, which is broken only by the linear friction term. The reason is that the smallness of $\langle v \omega^n \rangle$ for large-scale contributions implies the smallness of the respective correlation functions as well. This conclusion is supported by the data presented in Fig.\ref{fig:4}.

Let us show now that the mean profile can be obtained by analysis based on the conservation laws. The Navier-Stokes equation (\ref{Navier-Stokes}) itself is the momentum conservation law. Averaging it and taking the radial component, one obtains
 \begin{eqnarray}
 \partial_r\langle r v^2\rangle
 +r\partial_r \langle p \rangle
 =U^2 + \langle u^2\rangle.
 \label{Ber1}
 \end{eqnarray}
In deriving (\ref{Ber1}) we exploited isotropy and the incompressibility condition $\partial_\varphi u +\partial_r (r v)=0$. Taking the polar component of the averaged equation of (\ref{Navier-Stokes}), one finds
 \begin{eqnarray}
 r^{-1}\partial_r\left(r^2 \langle uv\rangle\right)
 =-\alpha r U,
 \label{v0}
 \end{eqnarray}
where again we exploited isotropy and incompressibility.

The left side of Eq.~(\ref{v0}) is the divergence of the flux of the averaged angular momentum $rU$. Thus the quantity $r\langle uv\rangle$ is the mean angular momentum flux. When $\langle uv\rangle$ is nonzero, the flow is irreversible, i.e. the sign of $\langle uv\rangle$ does not change upon the transformation $t\to -t$ while the sign of $U$ does. If $\langle uv\rangle$ does not decay faster than $r^{-2}$, then the sign of $\langle uv\rangle$ is opposite to that of $U$. Opposite signs of $U$ and $\langle uv\rangle$ imply that the momentum flows towards the vortex center (this is natural since the mean angular momentum density $rU$ {\it decreases towards the center}).

We now turn our attention to the energy balance equation. By taking a scalar
product of ${\bm v}$ with the Navier-Stokes equation (\ref{Navier-Stokes}) and
averaging, one gets the total energy density (containing both the mean flow and
fluctuations):
 \begin{eqnarray}
 {1\over r}\partial_r \left[rU\langle uv\rangle
 +r\left\langle\! v\!\left(\!{u^2+v^2\over2}+p\right)\!\right\rangle \right]
 \nonumber \\
 +\alpha\left(U^2+\left\langle u^2
 +v^2\right\rangle\right)
 =\langle {\bm f}\cdot{\bm v}\rangle.
 \label{En0}
 \end{eqnarray}
In deriving (\ref{En0}) we have neglected, again, viscosity. Indeed, viscosity mainly influences the direct cascade, dissipating enstrophy (squared vorticity), whilst energy dissipation by viscosity can be neglected \cite{BE12} (in the numerics we use hyper-viscous dissipation).

We now consider the internal region of the vortex, where $u,v\ll U$, see Figs. \ref{fig:5} and \ref{fig:6}, which demonstrate that fluctuations inside the coherent vortex are suppressed in comparison to the mean flow. It is a consequence of the large value of the mean velocity gradient $\sim U/r$, growing toward the center of the vortex. The relative strength of fluctuations increases as $r$ grows and on the periphery where $r\simeq L$, fluctuations become of the order of the average flow. Considering the vortex interior, we neglect $\langle u^2+v^2\rangle$ in comparison to $U^2$, and also odd in $v$ terms since they contain two small parameters, related to the smallness of $\alpha$ and that of the fluctuations. Substituting $\langle {\bm f}\cdot{\bm v}\rangle =\epsilon$ one obtains
 \begin{eqnarray}
 \epsilon= {1\over r}\partial_r\left(r
 U\langle uv\rangle\right)+\alpha U^2.
 \label{En1}
 \end{eqnarray}
Note that the same approximation is made in considering logarithmic turbulent boundary layers \cite{LL}, but there $\epsilon$ is the energy dissipation rate, whose coordinate dependence is unknown {\it a priori}. In our case, $\epsilon$ is the pumping term independent of coordinates, which allows us to solve the problem.
Combining the two Eqs.~(\ref{v0}) and (\ref{En1}) we find an $r$-independent mean polar velocity
 \begin{equation}
 U^2=3\epsilon/\alpha ,
 \label{Uni}
 \end{equation}
which is in excellent agreement with the numerics, both in value and in the $r$-independence, see Fig.~\ref{fig:3}.

It follows from~\eqref{Uni} that the second term in (\ref{En1}) is equal to $2\epsilon$, i.e. at every point inside the vortex the energy transfer from outside brings twice more than the local inverse energy cascade. Substituting expression~\eqref{Uni} into Eq.~(\ref{Ber1}) and neglecting $u^2$ and $v^2$, in comparison to $U$, we obtain for the pressure
 \begin{equation}
 p(r)= (3\epsilon/\alpha)\ln (r/R),
 \label{pressure}
 \end{equation}
 where $R\sim L$.  We present the radial profile of the pressure around the vortex condensate in Fig.~\ref{fig:7}. One extracts from the numerical data $R/L = 0.143$, which is approximately the size of the coherent vortex, see Fig.~\ref{fig:2}.

To conclude, we developed a theoretical scheme describing the mean velocity profile inside the coherent vortices and showed that $U$ is $r$-independent. Within the vortex, we found that velocity fluctuations are suppressed. Towards the periphery, velocity fluctuations become comparable to the mean flow, both of which can be estimated as $\left(\epsilon/\alpha\right)^{1/2}$, in accordance with the balance between the energy production and large-scale dissipation. For small $r$, the profile $U=\mathrm{const}$ is correct down to the vortex core.

This work was carried out under the HPC-EUROPA2 project (project number: 228398) with the support of the European Commission Capacities Area - Research Infrastructures Initiative.  The work at the Weizmann Institute was supported by the grant of  the Minerva Foundation with funding from the German Ministry for Education and Research. The work in Russia was supported by the RFBR Grant No. 2-02-01141a.

\end{document}